\documentclass[sn-mathphys,Numbered]{sn-jnl}

\usepackage{graphicx}%
\usepackage{multirow}%
\usepackage{amsmath,amssymb,amsfonts}%
\usepackage{amsthm}%
\usepackage{mathrsfs}%
\usepackage[title]{appendix}%
\usepackage{xcolor}%
\usepackage{textcomp}%
\usepackage{manyfoot}%
\usepackage{booktabs}%
\usepackage{algorithm}%
\usepackage{algorithmicx}%
\usepackage{algpseudocode}%
\usepackage{listings}%

\begin{document}

\title[]{Proposal to Observe Transverse Sound in Normal Liquid $^3$He in Aerogel}


\author{\fnm{Priya} \sur{Sharma}}\email{ps0072@surrey.ac.uk}

\affil{\orgdiv{Advanced Technology Institute}, \orgname{University of Surrey}, \city{Guildford}, \postcode{GU1 7XH}, \state{Surrey}, \country{United Kingdom}}


\abstract{In the Fermi liquid metallic state, a static local magnetic moment is induced on the application of a circularly polarized electromagnetic wave, via the Inverse Faraday Effect(IFE). The direction of this moment is along the direction of propagation of light, and the magnitude of the moment depends on the frequency of  light, the temperature and various material parameters characteristic of the metal. I propose an analogous effect in the Fermi liquid state of $^3$He. A static circulating current is induced when liquid $^3$He is driven by a circularly polarized transverse acoustic wave. For liquid $^3$He filled into aerogel, the coupled system supports a low attenuation transverse sound mode. I  estimate the magnitude of induced circulating currents for this system and find that these are within the range of experimental measurement in the low attenuation regime.  The axis of circulation is along the direction of propagation of the acoustic wave. I propose this analogue of the inverse Faraday effect as a scheme to experimentally demonstrate the propagation of transverse sound in $^3$He-aerogel.}




\maketitle

\section{Introduction}\label{sec:Intro}

The propagation of transverse sound in pure liquid $^3$He was first predicted by Landau. While the existence of shear waves demonstrates the Fermi liquid nature of liquid $^3$He, shear waves have never been observed in the normal liquid state of $^3$He, primarily due to the large attenuation of this sound mode \cite{WolfleBook}. However, transverse sound and the acoustic Faraday effect have been observed in superfluid $^3$He \cite{NatureAcousticFaraday} confirming the existense of transverse sound, as predicted in this system.

$^3$He in aerogel is a realisation of a disordered Fermi liquid state. This system has a rich superfluid phase diagram including new superfluid phases unseen in the bulk and has been studied extensively over the past two decades \cite{HalperinReview}. The superfluid state realises a disordered $p$-wave order parameter and the effect of impurities, whose density and isotropy (or lack of it) can be controlled, has been a subject of continued interest in the context of unconventional superconductors and novel phenomena. In this paper, I focus on the Fermi liquid state of $^3$He in aerogel.

Acoustic studies have conventionally been a powerful probe of liquid $^3$He physics. For the $^3$He-aerogel system, the first to zero sound transition, characteristic of the Fermi liquid state, is not observed \cite{Nomura:2000,Gervais:2000}. This is explained by the collision drag effect \cite{DragForce:2001}, viz., coupling of simultaneous oscillations of the liquid $^3$He and the aerogel matrix. The effect of collision drag on transverse sound is discussed in \cite{HigashitaniPRL:2002} and the authors find that the attentuation in the $^3$He-aerogel coupled system exhibits a  different behaviour as a function of temperature, as opposed to the pure $^3$He case. At low temperatures, the attenuation remains finite due to impurity scattering( as opposed to attenuation going to zero in the pure liquid case). At higher temperatures, the attenuation crosses over to a regime in which hydrodynamic
transverse sound can propagate over a long distance with the aid of the elasticity of the aerogel. However, inspite of the existence of this low-attenuation mode, transverse sound has not been experimentally observed in $^3$He-aerogel. Although shear impedance measurements have been attempted in this system, no conclusive evidence of a shear mode could be made from the response of an acoustic transducer to  pressure modulation from a transverse sound mode in an acoustic cavity \cite{GervaisThesis}. In this paper, I present a proposal to  demonstrate the existence of propagating transverse sound in the $^3$He-aerogel system, by measuring static circulating currents induced when the $^3$He-aerogel system is subject to a circularly polarized shear wave. These circulating currents are induced ONLY for transverse acoustic perturbations driven by a circularly polarized drive, thus providing an unambiguous demonstration of the propagation of transverse sound in normal-state $^3$He-aerogel. The motivation for this proposal is derived from an analogous effect in the Fermi liquid state of metals, the Inverse Faraday Effect(IFE). 

I introduce the IFE in metals in Sec \ref{sec:IFEMetals}. The induced circulation of orbital currents of electrons (which can be expressed in terms of a local magnetic moment) can be calculated within the Eilenberger formulation of quasiclassical theory.  In Section \ref{sec:Analogue}, I make the analogy to the liquid $^3$He case and show that the estimate for the induced circulating current is large enough to detect in experiments.

\section{Inverse Faraday Effect in Metals}\label{sec:IFEMetals}

The Inverse Faraday Effect(IFE) was first proposed by Pitaevskii \cite{Pitaevskii:1960} as an effect of a time-varying electric field on the stress tensor in a dispersive medium. \cite{Malmstrom:1965} coined the name and first reported an observed induced magnetization in various liquids and glasses. The effect has since been observed in several magnetic materials \cite{RevModPhys:2010} and more recently, in metals \cite{NatPhotonics:2020}. Theoretical studies of the IFE have been reported using semiclassical \cite{Majedi:2021}, {\it{ab initio}}\cite{Oppeneer:2018} and Ginzburg Landau \cite{Buzdin:2022,BuzdinAVortices:2022} schemes in a variety of materials. A microscopic theory for the IFE using quasiclassical Green's functions has been recently developed and used to predict the induced magnetic moments in metals \cite{PSIFE}. The primary response of electrons in a metal exposed to circularly polarized electromagnetic radiation of frequency $\omega$, is  to the oscillating electric field given by ${\bf{E}} = {\bf{E}}_0 e^{{{i{\bf{K}}.{\bf{R}}}} - i\omega t}$. Ignoring the band structure of electrons and for an isotropic Fermi surface, the corrections to the Green's functions for electrons responding to the oscillating electric field, give a static induced current density via the IFE. The rotational part of this current density is the curl of a magnetic moment given by\cite{PSIFE}, 
\begin{equation}
\label{M-ind}
{\bf{M}}_{ind} = \mu_B \frac{\omega_p^2}{2\omega(\omega^2 + 16\Gamma^2)}\, (i\epsilon_0\,{\bf{E}}_0\times{\bf{E}}_0^{\star})\zeta^2(\frac{\omega}{2k_B T})\,\,\,,
\end{equation}
in the $\bf{K}\rightarrow 0$ limit. Here,
 $\Gamma$ is the rate at which quasiparticles scatter from impurities/disorder, $\omega_p$ is the plasma frequency in the metal and $\epsilon_0$ and $\mu_B$ are the free-space permittivity and the Bohr magneton, respectively. $k_B$ is the Boltzmann constant and $T$ is the ambient temperature. The function $\zeta^2(x) \leq 1$ over the temperature range and is discussed in \cite{PSIFE}.   The DC magnetization is induced via a nonlinear effect upon application of an AC external field and can be regarded as a "rectification" effect.

 \section{Analogue  in  $^3$He}\label{sec:Analogue}

 To make the analogy to the charge neutral Fermi liquid $^3$He, I follow closely along the lines of the Keldysh formulation ~\cite{Keldysh:1965} of quasiclassical theory ~\cite{Eilenberger:1968, LarkinOvchinnikov:1969} employed in \cite{PSIFE} to calculate the IFE in metals. I consider the response of $^3$He quasiparticles  to an oscillating shear field of an impinging transverse wave, ${\bf{U}} = {\bf{U}}_0 e^{i{\bf{q}}.{\bf{R}} - i\omega t}$. This is analogous to the oscillating electric field of light in the IFE. I estimate the second-order current response analogous to the induced current density via the IFE.

The quasiclassical Green's function, $\hat{g}$ is the propagator for quasiparticles with effective mass $m^{\star}$, energy $\varepsilon$ and Fermi momentum ${\bf{p}} = m^{\star} {{v}}_F \hat{p}$, given by solutions to the Eilenberger equation ~\cite{Eilenberger:1968}. Observables such as the quasiparticle density are calculated from the Keldysh components of the Green's function, $\hat{g}^K$, viz., $\rho = \int \frac{d\varepsilon}{2\pi} \frac{d^3 p}{(2\pi\hbar)^3} \mathscr{T} (\hat{g}^K)$; here, $\mathscr{T}$ refers to a trace over Nambu and spin indices. I expand $\hat{g}^K$  in the external field $\bf{U}$, $\hat{g}^K = \hat{g}_0^K + \hat{g}_1^K + \hat{g}_2^K$, with $\hat{g}_i^K \propto \mathcal{O}(\bf{U}^i)$ being the $i$-th order correction to $\hat{g}^K$ and evaluate the current response. The details of the definitions are in the Appendix. I am interested in the second-order $\mathcal{O}(\bf{U}^2)$ corrections to the current density, $\bf{j}_2$, averaged over the period of the oscillating field. It has been shown \cite{PSIFE} that  
\begin{equation}
\label{jIFE}
    {\bf{j}}_2 =  \langle \,\bar{\rho}_1({\bf{p}})\,\, \bar{\bf{v}}_1({\bf{p}}) \,\rangle_{{\bf{{p}}},(2\pi/\omega)}\,\,\,,
\end{equation}
where $\langle ...\rangle_{{\bf{p}},(2\pi/\omega)} \equiv (\frac{\omega}{2\pi})\int dt \int \frac{d^3 p}{(2\pi)^3} $ refers to an average over the Fermi surface and an average over a time-period of the field; $\bar{\rho}_1$ and $\bar{\bf{v}}_1$ are the first-order corrections to the quasiparticle density and velocity, respectively viz., 
\begin{eqnarray}
\label{bar-defn}
    \bar{\rho}_1({\bf{p}}) &\equiv& N_f \int d\varepsilon\, \mathscr{T} \hat{g}_1^K({\bf{p}},\varepsilon, t) \\
    \nonumber 
    \bar{\bf{v}}_1({\bf{p}})  &\equiv&   \frac{N_f}{\rho_0}{\bf{v}}_F \int {d\varepsilon'} \mathscr{T}\hat{\tau}_3 \hat{g}_1^K({\bf{p}},\varepsilon', t) \,\,\,,
\end{eqnarray}
where $\hat{\tau}$ are Pauli matrices in Nambu space, $\rho_0$ is the equilibrium density of quasiparticles, $N_f$ is the density of states at the Fermi level and $\mathscr{T}$ refers to the trace over the Nambu and spin degrees of freedom.
The first-order corrections $\hat{g}_1^K$ can be obtained by solving the Eilenberger equations. The Keldysh functions $\hat{g}^K$ may also be expressed in terms of quasiparticle distribution functions, $\hat{n}$ and the Retarded and Advanced functions,
\begin{equation}
    \hat{g}^K = -2 (\hat{g}^R\circ\hat{n} - \hat{n}\circ\hat{g}^A)\,\,\,.
\end{equation}
The distribution function then satisfies the Landau-Boltzmann transport equation with driving terms given by the collision integrals for quasiparticles scattering via various scattering channels.
The distribution functions may also be expanded in the external field as the quasiclassical propagators. Then the first order corrections may be related,
\begin{equation}
\label{g1K-n}
    \hat{g}_1^K = -2 (\hat{g}_1^R\circ\hat{n}_0 - \hat{n}_0\circ\hat{g}_1^A) -2 (\hat{g}_0^R\circ\hat{n}_1 - \hat{n}_1\circ\hat{g}_0^A)\,\,\,.
\end{equation}
Solving the Eilenberger equation for $\hat{g}_1^{R,A}$ gives $\hat{g}_1^{R,A} = 0$, as shown in the Appendix. Then equation(\ref{g1K-n}) reduces to
\begin{equation}
\label{g1n1}
    \hat{g}_1^K =-2 (\hat{g}_0^R\circ\hat{n}_1 - \hat{n}_1\circ\hat{g}_0^A)\,\,\,.
\end{equation}

\cite{HigashitaniPRL:2002} have calculated the first-order corrections to the quasiparticle distribution function by solving the Boltzmann-Landau transport equation for $^3$He-aerogel, including elastic scattering of quasiparticles from the aerogel matrix, the motion of the aerogel driven by the acoustic field, and mutual collisions between quasiparticles that provide an inelastic scattering channel. They find that the corrections to the distribution function $ n_1({\bf{p}},\varepsilon) = f'(\varepsilon_{\bf{p}}^0)\nu_{\bf{p}}e^{i{\bf{q}}.{\bf{R}} - i\omega t}$, where $f'(\varepsilon_{\bf{p}}^0)$ is the equilibrium distribution function and $\nu_{\bf{p}} = \Sigma_{l=1}^2 \nu_l Y_l^1(\theta,\phi)$; $\theta$ and $\phi$ being the polar and azimuthal angles of the quasiparticle momentum $\bf{p}$ with respect to the direction of $\bf{q}$ respectively. The coefficients, $\nu_l$ depend on the Fermi liquid parameters in  $^3$He, the characteristic amplitude, frequency and wavelength of the driving sound wave and contain information about scattering parameters, including the relaxation times for impurity and mutual quasiparticle scattering. They represent the so-called collision drag effect due to the aerogel on the distribution function of quasiparticles, in response to an impinging transverse sound wave.

Using the result for $n_1$ from \cite{HigashitaniPRL:2002} in equation(\ref{g1n1}) and the notation $\hat{g}^X \equiv \Big( \begin{array}{c c} g^X & f^X\\ \underline{f}^X & \underline{g}^X\end{array}\Big)$,
\begin{equation}
\label{g1K}
    {g}_1^K = i\beta \sqrt{\frac{3}{2\pi}}sin\theta\,e^{i\phi}(\nu_1   + \nu_2\sqrt{5} \, cos\theta)\, sech^2 (\beta\varepsilon)\,\, e^{i{\bf{q}}.{\bf{R}} - i\omega t}\,\,\,\,\,,
\end{equation}
where $\beta = (2 k_B T)^{-1}$ and the equilibrium distribution function is given by the Fermi function.
I can now use equation(\ref{g1K}) in equation(\ref{jIFE}-\ref{bar-defn}) to find $\bf{j}_2$, which represents the static current response to an applied oscillating field analogous to the "rectification" represented by the IFE. 

The objective is to calculate the time-averaged corrections to the current density ${\bf{j}}$ that are of second order in the field, ${\bf{j}}_2 \propto \mathcal{O}({\bf{U}}^2)$,
\begin{equation}
\label{j2}
{\bf{j}}_2 =  \langle \rho_2 {\bf{v}}_0 \rangle_{{\bf{p}},(2\pi/\omega)} +   \langle \rho_0 {\bf{v}}_2 \rangle_{{\bf{p}},(2\pi/\omega)} +   \langle \rho_1 {\bf{v}}_1 \rangle_{{\bf{p}},(2\pi/\omega)} \,\,\,\,\,.
\end{equation}
If I ignore the collision drag effect, $\rho_2({\bf{p}})$ is given by equation(\ref{bar-defn}) using the second-order corrections $\hat{g}_2^K$. $\hat{g}_2^K$ is obtained from equation(\ref{RAKeqns}) with the driving term $\propto ({\bf{v}}_F\cdot{\bf{q}})\,\hat{g}_1^K$. As $\hat{g}_1^K \propto ({\bf{v}}_F\cdot{\bf{q}})$ (analogous to the IFE case, ignoring collision drag), this gives $\langle \rho_2 {\bf{v}}_0 \rangle_{\bf{p}} \propto \langle {\bf{v}}_F({\bf{v}}_F\cdot{\bf{q}})({\bf{v}}_F\cdot{\bf{q}}) \rangle_{\bf{p}} = 0$. Similarly, the second term in equation(\ref{j2}) also gives a zero contribution. This leaves only the third term non-zero in ${\bf{j}}_2$ above. $\rho_1$ and ${\bf{v}}_1$ are given by $\hat{g}_1^K$, as in equation(\ref{bar-defn}). Ignoring the effects of collision drag beyond linear order  $\mathcal{O}({\bf{U}})$, ${\bf{j}}_2$ is given by equation(\ref{jIFE}-\ref{bar-defn}) with $\hat{g}_1^K$ given by equation(\ref{g1K})\cite{Footnote}.

Further, the dynamics of quasiparticles in $^3$He-aerogel is in the hydrodynamic limit over a large frequency range \cite{HigashitaniPRL:2002}. Assuming local equilibrium, I use the gradient representation derived in the Appendix and evaluate $\bar\rho_1$ and ${\bar{\bf{v}}}_1$,
\begin{eqnarray}
    \bar\rho_1 &=& \frac{N_f}{4}  \sqrt{\frac{3}{2\pi}} sin\theta\,\Big((\nu_1   + \nu_2\sqrt{5} \, cos\theta)\,\,\, e^{i\phi}e^{i{\bf{q}}.{\bf{R}} - i\omega t} - c.c.\Big)\\
    \nonumber
    \bar{\bf{v}}_1 &=& -\frac{N_f {\bf{v}}_F}{4\rho_0} \,\,(\frac{i{\bf{v}}_F\cdot\nabla}{\omega} )\sqrt{\frac{3}{2\pi}} sin\theta\,\Big((\nu_1   + \nu_2\sqrt{5} \, cos\theta)\,\,\, e^{i\phi}e^{i{\bf{q}}.{\bf{R}} - i\omega t} + c.c.\Big)\,\,\,.
\end{eqnarray}
To estimate the magnitude of the induced current, I estimate the magnitude of $\nu_1 \propto \nu_{1a}({\bf{v}}_F\cdot{\bf{q}}){\bf{U}}_0$. Here ${\bf{U}}_0$ is the amplitude of the shear wave which is circularly polarized, i.e., ${\bf{U}}_0 \neq {\bf{U}}_0^\star$; and $\nu_{1a}$ contains details about aerogel parameters including quasiparticle relaxation times and the geometric mean free path of the aerogel. ${\bf{j}}_2$ has both rotational and irrotational terms. The irrotational part corresponds to currents in the plane of polarization of the transverse acoustic drive. Writing out the rotational part of the current, I estimate the time-averaged $({\bf{j}}_2)_{rot}$ thus,
\begin{eqnarray}
    ({\bf{j}}_2)_{rot} &\propto& \frac{3i\,N_f^2 v_F^2}{16 \pi \omega \rho_0} \, \nu_{1a}^2 v_F^2 q^2 \,\,\nabla\times ({\bf{U}}_0 \times {\bf{U}}_0^\star)\,\,\,\\
    \nonumber
    &\sim& \frac{3 i k_F \,\nu_{1a}^2}{16 \pi^3 \omega} (\frac{ v_F^2 q^2}{\hbar^2}) \nabla\times ({\bf{U}}_0 \times {\bf{U}}_0^\star)\,\,\,.
\end{eqnarray}
For a shear wave of frequency $\omega = 20 MHz$ and $q \sim 10^6 m^{-1}$, used in \cite{HigashitaniPRL:2002}, and for an aerogel sample of size $\sim 1 cm$ on each side, I estimate the magnitude of this current density to be $\sim 10^{-6} kg \,m^{-2} sec^{-1}$. The curl in the expression above reveals the circulating nature of currents. Clearly, this vanishes for linearly polarized drive where ${\bf{U}}_0 = {\bf{U}}_0^\star$. The currents circulate around an axis along the direction of ${\bf{U}}_0 \times {\bf{U}}_0^\star$. For a circularly polarized shear wave impinging along the axis of a cm-sized cylindrical sample of aerogel, the currents circulate about an axis that is parallel (or antiparallel) to the axis of the cylinder, the specific direction (parallel or antiparallel) being determined by the handedness of polarization of the shear wave. A schematic is shown in Fig. 1. I estimate the total current for a $1 cm$ tall sample to be $\sim  10^{-10} kg/sec$, which is within the range of mass current measurement by experiment\cite{RMPJosephson:2002}.

\begin{figure}
    \label{fig:Schematic}
    \includegraphics[width = 0.9\textwidth]{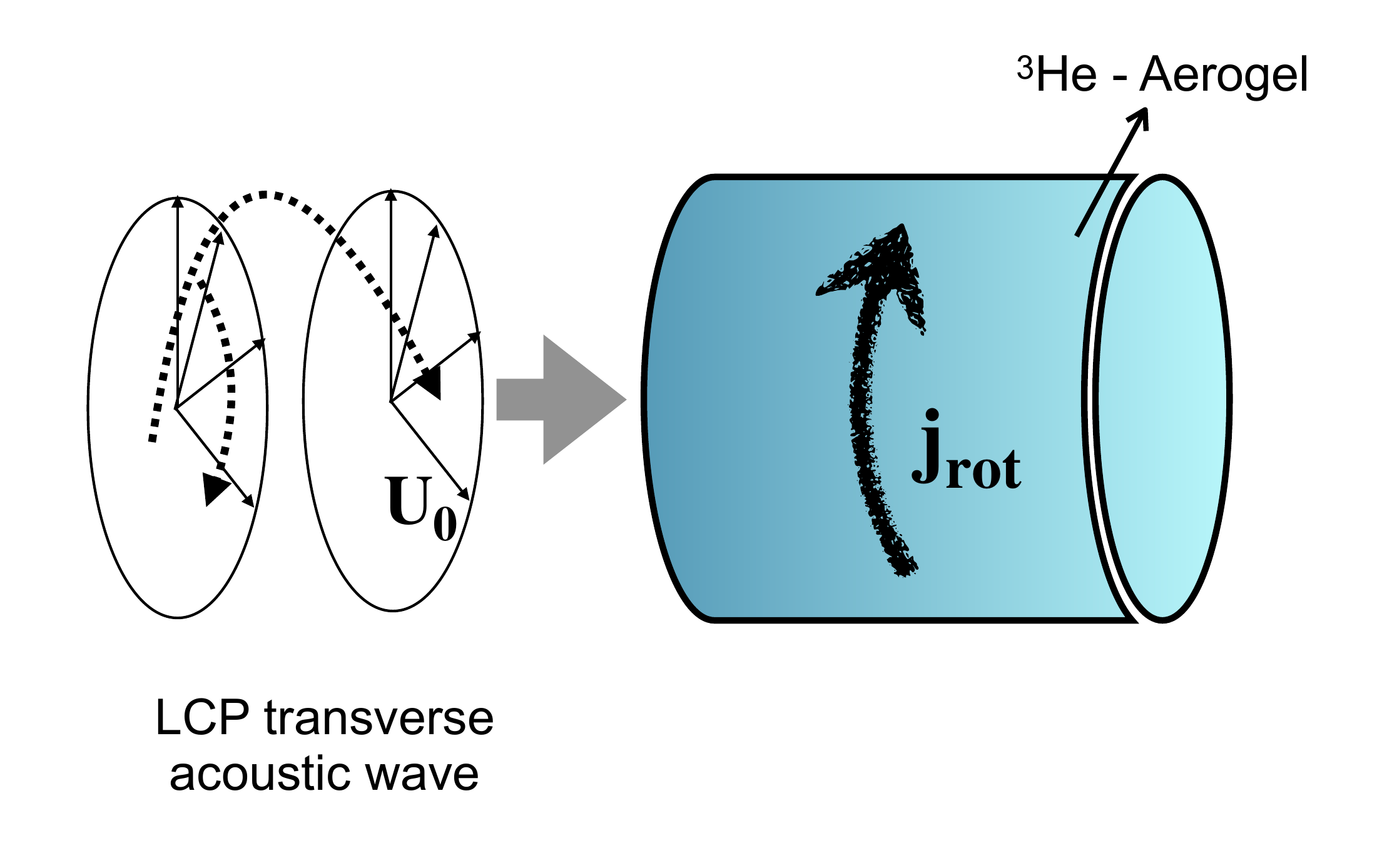}
    \caption{An illustration of circulating currents ${\bf{j}}_{rot}\propto U_0^2$ induced by an impinging left circularly polarized (LCP) transverse acoustic wave of amplitude ${\bf{U}}_0$. The direction of the induced current will be reversed for right circularly polarized drive.}
\end{figure}

The collision drag effect couples simultaneous oscillations of liquid $^3$He with that of the aerogel matrix leading to a decrease in the attenuation of transverse sound at higher temperatures \cite{HigashitaniPRL:2002}. This decrease in attenuation depends on the velocity of transverse sound (not reported thus far) and can be up to two orders of magnitude for higher sound velocities, for temperatures  $\sim 50-100 mK$ \cite{HigashitaniPRL:2002}. Thus, in this temperature range, the aerogel supports the propagation of transverse sound over longer distances enabling the detection of circulating currents produced as a result. While these currents would also exist in the bulk via the mechanisms described in this work, the attenuation of transverse sound in pure liquid $^3$He is rather high (of the order of $500 cm^{-1}$) rendering the induced currents weak and undetectable. 

Linearly polarized transverse acoustic waves have been  successfully realised experimentally and used in acoustic studies of liquid $^3$He \cite{WolfleBook,Nomura:2000}. Circularly polarized shear waves have, however, not been favourably generated experimentally for such studies yet. The generation of circularly polarized shear waves for applications ranging from solid-state to geophysical are based on the use of two transverse transducers that generate linearly polarized shear waves in orthogonal directions, locked in phase  to generate a resultant circularly polarized wave. While a perfect circularly polarized shear wave might be difficult to generate due to intrinsic differences between transducers and (lack of) precision of phase locking \cite{Man-email}, a "good" circularly polarized wave is not necessary to observe the induced current predicted in this work. An elliptically polarized shear wave or in particular, any shear drive, ${\bf{U}}$ with ${\bf{U}} \times {\bf{U}}^\star \neq 0$ is sufficient to induce circulating currents described above (${\bf{U}} \times {\bf{U}}^\star = 0$ for linearly polarized shear drive).

\section{Summary}

I consider the effect in normal-state Fermi liquid $^3$He analogous to the IFE in metals. The response of $^3$He quasiparticles to an impinging circularly polarized shear wave is to transfer the angular momentum of the wave to quasiparticles resulting in an induced circulating current. As $^3$He in aerogel is predicted to support a low-attenuation transverse sound mode,  I estimate the magnitude of the induced circulating current for $^3$He-aerogel. For high porosity aerogels, I estimate this current to be within the range of experimental measurement for sound waves in the low-attenuation $MHz$ frequency range for cm-sized samples. As this circulating current is induces only for shear drives, I propose the observation of this current as a demonstration of transverse sound propagation in liquid $^3$He-aerogel.






\backmatter

\bmhead{Acknowledgments}
I thank Alexander V. Balatsky for introducing me to the Inverse Faraday Effect. I am grateful for useful discussions with Johannes Pollanen and Man Nguyen. 
I acknowledge funding from the Daphne Jackson Trust, UK and the Engineering and Physical Sciences Research Council, UK.






\begin{appendices}
\section{Appendix}

The central object of the theory is the quasiclassical Green's function, $\check{g}$ which is the propagator for quasiparticles of effective mass $m^{\star}$ with energy $\varepsilon$ and Fermi momentum ${\bf{p}} = m^{\star} {{v}}_F \hat{p}$, given by the solutions to the equation,
\begin{equation}
\label{Kitaeqn}
[\varepsilon\check{\tau}_3 - \check{\sigma},\check{g}]_{\circ} + i{\bf{v}}_F\cdot\partial_{\vec{R}}\check{g} 
= \check{0} \,\,\,,
\end{equation}
where 
\begin{equation}
\label{gNambu}
\check{g} = (\begin{array}{c c} \hat{g}^R & \hat{g}^K\\ \hat{0} & \hat{g}^A \end{array}) \,\,\,;\,\,\, \check{\tau}_3 = (\begin{array}{c c} \hat{\tau}_3 & \hat{0}\\ \hat{0} & \hat{\tau}_3 \end{array})\,\,\,;\,\,\,\check{0}= (\begin{array}{c c} \hat{0} & \hat{0}\\ \hat{0} & \hat{0} \end{array})\,\,\,.
\end{equation}
Here, $\hat{g}^{R,A,K}$ are the Retarded, Advanced and Keldysh propagators and the $\hat{\,}$ refers to matrices in Nambu space. $\hat{\tau}_{j}$ $(j = 1,2,3)$ are the Pauli matrices in Nambu space. $\check{\sigma}$ is the self-energy with corresponding Retarded, Advanced and Keldysh components. 
The circle product is defined as :
\begin{equation}
\label{circdefnew}
    \hat{a}\circ\hat{b}({\bf{p}},{\bf{R}},\varepsilon,t) = \hat{a}({\bf{p}},{\bf{R}},\varepsilon - \frac{1}{2i} \frac{\partial}{\partial t_2}, t_1) \, \hat{b}({\bf{p}},{\bf{R}}, \varepsilon + \frac{1}{2i} \frac{\partial}{\partial t_1}, t_2)  \mid_{t_1 = t_2 = t}\,\,\,,
\end{equation}
where the $\partial/\partial t_1$ should be understood as acting on the left ~\cite{RainerSerene:1983}.
 This formulation is applicable for arbitrary external frequencies, $\omega$ much smaller than the Fermi energy, $\omega\ll\varepsilon_F$. Putting equation(\ref{gNambu}) in equation(\ref{Kitaeqn}) yields separate equations for the $R,A,K$ propagators,
\begin{eqnarray}
\label{RAKeqns}
[\varepsilon\hat{\tau}_3 - \hat{\sigma}^{R,A},\hat{g}^{R,A}]_{\circ} + i{\bf{v}}_F\cdot\nabla\hat{g}^{R,A} 
&=& \hat{0}\,\,\,;\\
\nonumber
(\varepsilon\hat{\tau}_3 - \hat{\sigma}^R)\circ\hat{g}^K - \hat{\sigma}^K\circ\hat{g}^A + \hat{g}^R\circ\hat{\sigma}^K - \hat{g}^K\circ(\varepsilon\hat{\tau}_3 - \hat{\sigma}^A) +  i{\bf{v}}_F\cdot\nabla\hat{g}^K 
&=& \hat{0} \,\,\,.
\end{eqnarray}

Consider an oscillating shear field of frequency $\omega$ and wave-vector $\bf{q}$ applied to the $^3$He-aerogel system, ${\bf{U}} = {\bf{U}}_0 e^{i{\bf{q}}.{\bf{R}} - i\omega t}$. The response of quasiparticles to this field can be calculated using equations(\ref{RAKeqns}). Expand $\hat{g}$  in the external field ${\bf{U}}$,
\begin{equation}
\label{gexp}
\hat{g}^X = \hat{g}_0^X + \hat{g}_1^X + \hat{g}_2^X\,\,\,\,\,,(X = R,A,K)
\end{equation}
with $\hat{g}_i^X \propto \mathcal{O}({\bf{U}}^i)$ being the $i$-th order correction to $\hat{g}^X$ and evaluate the current response. The current density operator at a space-time point $({\bf{R}},t)$ is given by 
\begin{equation}
\label{j-defn}
{\bf{j}}({\bf{R}},t) =  \,\rho({\bf{R}},t)\, {\bf{v}}({\bf{R}},t) \,\,\,,
\end{equation}
where $\rho({\bf{R}},t)$ and ${\bf{v}}({\bf{R}},t)$ are the local density and velocity of quasiparticles, respectively. (${\bf{R}}$ and $t$ are the centre of mass space and time coordinates).
\begin{eqnarray}
\label{nv}
    \rho({\bf{R}},t) &=&  \int \frac{d^3 p}{(2\pi)^3} \rho({\bf{p}},t) e^{i{\bf{p}}\cdot{\bf{R}} }\\
    \nonumber 
    {\bf{v}}({\bf{R}},t) &=&    \int \frac{d^3 p'}{(2\pi)^3} {\bf{v}}({\bf{p}}\,',t) e^{i{\bf{p}}\,'\cdot{\bf{R}}}
\end{eqnarray}
Using equations(\ref{nv}) in equation(\ref{j-defn}),  I get ${\bf{j}}$,
\begin{equation}
\label{j2n1v1}
    {\bf{j}}({\bf{R}},t) =  \langle \,\bar{\rho}({\bf{p}})\,\, \bar{v}({\bf{p}}) \,\rangle_{\bf{p}}\,\,\,,
\end{equation}
where $\langle ...\rangle_{\bf{p}} \equiv \int \frac{d^3 p}{(2\pi)^3}$ refers to an average over the Fermi surface and $\bar{\rho}$ and $\bar{\bf{v}}$ are defined in equation(\ref{bar-defn}).

At equilibrium, $\hat{g}_0$ is given by the solution to the Eilenberger equation in the absence of external fields,
\begin{equation}
\label{g0}
    \hat{g}_0^{R,A} = \pm i \, \hat{\tau}_3\,\,\,,\,\,\,\hat{g}_0^K =  - 2 i\, tanh(\beta\varepsilon) \,\hat{\tau}_3\,\,\,,
\end{equation}
where $\beta\equiv (2 k_B T)^{-1}$ and $\varepsilon^{R,A} = \varepsilon \pm i0+$. 
I use the notation for the Nambu elements, $\hat{g}^X \equiv \Big( \begin{array}{c c} g^X & f^X\\ \underline{f}^X & \underline{g}^X\end{array}\Big)$. 
$g^X$ and $\underline{g}^X$ (and analogously $f^X$ and $\underline{f}^X$) are related by symmetries, $[\hat{g}^R]^{\dagger} = -\hat{\tau}_3\hat{g}^A\hat{\tau}_3$ and $[\hat{g}^K]^{\dagger} = \hat{\tau}_3\hat{g}^K\hat{\tau}_3$.
$\hat{g}_1^X$ are given by solutions to equations(\ref{RAKeqns}). Using the notation in equation(\ref{gexp}) and the definition(\ref{circdefnew}), to $\mathcal{O}({\bf{U}})$,
\begin{eqnarray}
\label{MatrixEqnsgRAK}
    (\varepsilon - \sigma^R - \frac{1}{2i}\frac{\partial}{\partial t})\,\hat{\tau}_3\hat{g}_1^R - (\varepsilon - \sigma^R &+& \frac{1}{2i}\frac{\partial}{\partial t})\,\hat{g}_1^R\hat{\tau}_3 + i{\bf{v}}_F\cdot\nabla \hat{g}_1^R   = 0\,\,\,.
\end{eqnarray}
Examining these equations,  I seek solutions of the form
\begin{equation}
\label{g-ansatz}
    \hat{g}_1^X = \hat{g}_+^X\,e^{i\omega t} + \hat{g}_-^X\,e^{-i\omega t}\,\,\,.
\end{equation}
$\hat{g}_\pm^X = (\begin{array}{c c} g_\pm^X & 0\\ 0 & \underline{g}^X_\pm \end{array})$, using the notation defined above, with the symmetry $\underline{g}^R(\hat{p},\varepsilon) = -g^R(-\hat{p},-\varepsilon)^{\star}$ and $\underline{g}^K(\hat{p},\varepsilon) = g^K(-\hat{p},-\varepsilon)^{\star}$. 
Using the ansatz, equation(\ref{g-ansatz}) in equation(\ref{MatrixEqnsgRAK}), 
I get equations for $g_\pm^R$ ($\underline{g}_\pm$ can be obtained using symmetries),
\begin{eqnarray}
\label{gR}
    (\varepsilon - i\Gamma \pm \frac{\omega}{2})\,g_\mp^R - (\varepsilon - i\Gamma \mp \frac{\omega}{2})\,g_\mp^R + i{\bf{v}}_F\cdot\nabla g_\mp^R  
    &=& 0\,\,\,\\
    \nonumber
    \Rightarrow g_\pm^{R} = \underline{g}_\mp^R = g_1^R &=& 0\,\,\,.
\end{eqnarray}
Similarly, $\hat{g}_1^A = 0$.

 Assuming local equilibrium, to lowest order in spatial gradients and to zeroth-order in applied fields, the Eilenberger equation (\ref{RAKeqns}) for $\hat{g}^K$ may be written as
\begin{eqnarray}
\label{grads}
    \varepsilon\circ\hat{g}^K - \hat{g}^K\circ\varepsilon + i {\bf{v}}_F\cdot\nabla\hat{g}^K &=& 0 \\
    \nonumber
    \hat{g}^K &=& - \frac{i{\bf{v}}_F\cdot\nabla\hat{g}^K}{\omega}\,\,\,.
\end{eqnarray}
I use  this gradient representation (to zeroth order in applied field) in equation(\ref{bar-defn}) to reveal the circulating nature of the induced current density.

\end{appendices}

\bibliography{sn-bibliography}

\end{document}